\documentclass[onecolumn,aps,floats,superscriptaddress,showpacs]{revtex4}
\usepackage{amsmath}

\usepackage{epsfig}
\usepackage{graphicx}
\usepackage{dcolumn}
\usepackage{bm}

\newcommand{\beq}{\begin{equation}}
\newcommand{\eeq}{\end{equation}}
\newcommand{\beqa}{\begin{eqnarray}}
\newcommand{\eeqa}{\end{eqnarray}}

\def\gapp{\lower.35em\hbox{$\stackrel{\textstyle>}{\sim}$}}
\def\lapp{\lower.35em\hbox{$\stackrel{\textstyle<}{\sim}$}}

\begin{document}
\bibliographystyle{apsrev}

\title{Effect of Coulomb interactions on the physical
observables of graphene}

\author{Mar\'{\i}a A. H. Vozmediano and F. Guinea}

\affiliation{Instituto de Ciencia de Materiales de Madrid,\\
CSIC, Cantoblanco, E-28049 Madrid, Spain.}

\date{\today}
\begin{abstract}
We give an update of the situation concerning the effect of electron--electron interactions
on the physics of the neutral graphene system at low energies.
We revise   old renormalization group results and the use of the 1/N expansion to address the questions of the possible opening of a low energy gap, and the magnitude of the graphene fine structure constant.  We emphasize the role of the Fermi velocity as the only free parameter determining the transport and electronic properties of the graphene system and revise its renormalization by Coulomb interactions on the light of  recent experimental evidences.

\end{abstract}
%
%
%
%
\maketitle
\section{Introduction}
\label{intro}
Graphene is considered to be a bridge between quantum field theory (QFT) and condensed matter physics due to the low energy description of the elementary excitations as massless Dirac fermions in two spacial dimensions. This description realizes some important toy models  used to study quark confinement and chiral symmetry breaking in QFT, particularly Quantum Electrodynamics in two spacial dimensions QED(3) \cite{S62,P84}. On the condensed matter point the QFT modeling of graphene raises two important questions: under a fundamental point of view it is not guaranteed that the neutral graphene system with Coulomb interaction is  a Fermi liquid \cite{L57}. On a more practical side, the QFT model has logarithmic divergences that need to be renormalized what introduces some subtleties in  fixing the value of the graphene fine structure constant affecting the experiments.

Electron--electron interactions were
considered to be very small or even absent in the first experiments
\cite{Netal04,Netal05,Zetal05,Jetal07,BOetal07,ZSetal08,LLA09,NBetal08,Letal08} but
the recent advances in the synthesis and the big improvement in the quality of the samples have changed this view. The observation of the fractional Quantum Hall effect \cite{DSetal09,BGetal09}
and the experiments probing the zero Landau level in high magnetic fields \cite{GGetal11}
have renewed the interest on the role of Coulomb interactions on the intrinsic properties of the
system. Since the role of many body corrections to the physics
of graphene has been extensively studied in the literature (for a recent review see \cite{KUetal11} and the references inside),
we will here pick up these aspects that in our opinion are more interesting or less well understood. Here we just note that, since all other interactions are irrelevant at sufficiently low energies \cite{GGV94,V11}, the Coulomb interaction is the only feature affecting the physical properties of the clean suspended samples.

The low energy effective model for graphene being a QFT model suffers from one typical problem: perturbation theory calculations of any observable quantity are plagued by  infinities or spurious cutoff dependences. The standard QFT renormalization program  allows us to address two very important questions in the physics of graphene:
what is the strength of the Coulomb interaction at low energies - i. e. what is the actual value of the graphene fine structure constant --, and will the interactions lead to a band gap opening? We summarize the contents of this review  by giving answers to these questions, the second first: A gap will not open in graphene at low energies for moderate values  of the graphene fine structure constant $\alpha_G$. And: the infrared value of $\alpha_G$ is smaller than the assumed value based on a Fermi velocity of the order of $c/300$.

Note however that interactions are very much enhanced in the presence of strong magnetic fields. In this situation a gap can open at the neutrality point \cite{CLO08}. This article describes the physical low energy properties of neutral, clean graphene at zero magnetic field.

\section{ Graphene versus QED}
\label{qed}

\subsection{The continuum model}
One of the facts that has allowed the fast development in the understanding of the electronic properties of graphene is that there is a well established model Hamiltonian universally accepted by the community. This is another
similarity  with the physics of elementary particles. The standard non-interacting model for the electronic excitations around  a single Fermi point in graphene   (in units $\hbar=1$) is given by
\begin{equation}
{\cal H}=  v_F \int d^2 {\bf r} \bar{\psi}({\bf r})
\gamma^i\partial_i \psi ({\bf r})\;, \label{freeH}
\end{equation}
where $i=1,2$, $\bar\psi({\bf r})=\psi^+({\bf r})\gamma^0$, and
the gamma matrices can be chosen as $\gamma_x=\sigma_2,
\gamma_y=-\sigma_1, \gamma^0=\sigma_3 $ . $\sigma_i$ are the Pauli matrices and
$v_{F}$ is the Fermi velocity, the only free parameter in the Hamiltonian. When promoting the Hamiltonian to a QFT action we  have  the electron wave function as an additional parameter defining the theory.

And the electron-electron interactions are described by the Hamiltonian
\beq
{\cal H}_{\rm int}= e^2\int d^2 {\bf r}d^2 {\bf r'}
\frac{\psi^+({\bf r})\psi({\bf r})\psi^+({\bf r'})\psi({\bf r'})}{\vert {\bf r}-{\bf r'}\vert}
\label{CMcoulomb}.
\eeq
It is immediately seen that, unlike what happens in the usual two dimensional electron gas, the ratio of the Coulomb interactions and the kinetic energy in this system is a constant independent of the density and given by $g\sim\frac{e^2}{v_F}$  often taken as the graphene fine structure constant.

Since the model Hamiltonian is the relativistic Dirac Hamiltonian it saves a lot of efforts to adopt a covariant formulation and profit of the results already known in QFT. The fact that the numerical value of Fermi velocity does not equal the speed of light  is not important, the model is still relativistic with a different light cone. What is very different and interesting under a QFT point of view is the fact that the velocity defining the light cone is renormalized by the interactions.

Following the quantum field theory nature of the model, we replace the instantaneous Coulomb interaction of eq. (\ref{CMcoulomb})  by a local gauge interaction through a minimal coupling:
$$L_{\rm int}=g\int d^2x dt j^\mu (x,t)A_\mu (x,t)\;, $$
where the electron current is defined as
$$j^\mu =(\overline\Psi\gamma^0\Psi,v_F \overline\Psi\gamma^i\Psi)\;.$$
With this substitution we have formally an effective Lagrangean
\beq
{\cal L}= \int d^2 {\bf r} \; d t \; \bar{\psi}
\; \gamma^\mu[\partial_\mu + i e A_\mu] \psi \;,
\label{fullL}
\eeq
very similar to that of massless QED in two spatial dimensions. What makes the graphene model different from QED in any number of dimensions is the gauge boson propagator. As discussed at length in \cite{GGV94,JGV10,V11}, a gauge field propagator in QFT has a $1/k^2$ dependence while the effective Coulomb interaction in graphene propagates as $1/k$. This is due to the fact that only the charges are confined to live in the two dimensional surface defined by the sample. The photons propagate in the three dimensional space.  This is a crucial aspect that makes the model interesting and behaving more like the scale invariant massless QED(3+1) rather than the super--renormalizable QED(2+1). The renormalization of the Fermi velocity is a consequence of this fact.

\subsection{Perturbative renormalization}
\begin{figure}
\begin{center}
\includegraphics[scale=0.5]{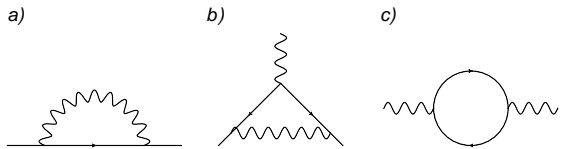}
\caption{Primitively divergent Feynman graphs in QED(4). Fermion self--energy (a),  vertex (b) and photon self-energy (c).}
\label{primitive}
\end{center}
\end{figure}

In standard QED(4) there are three one loop diagrams that have logarithmic singularities (see Fig. \ref{primitive}) and three free parameters in the theory (the coupling constant and the electron and photon wave functions).
The theory is strictly renormalizable in the sense that all divergences at any order in perturbation theory can be cured by a proper redefinition of the parameters. Subsequently the physical values of these parameters are fixed by experimental data (renormalization conditions). On the other hand, QED(3) is a super-renormalizable theory meaning that the coupling constant has dimensions of energy what improves the convergence of the perturbative series. It has less divergences and these can be properly renormalized. In fact, massless QED (2+1) is ultraviolet finite although it has infrared divergences \cite{MRS05}. Graphene sits in between  QED(3) and QED(4).  Of the three graphs shown in Fig. \ref{primitive} only the electron self--energy diverges and in the case of considering a static photon propagator only the spatial part has a logarithmic divergence.
In the static model, the electron and photon propagators in momentum space are given by
\begin{equation}
G_0(k^0, {\bf k})=i\frac{\gamma^0 k_0+ v\bm{\gamma}
\cdot {\bf k}}{-(k^0)^2+v^2{\bf k}^2}\;,
\label{elprop}
\end{equation}
\begin{equation}
\label{barephoton}
\Pi_0( {\bf k})=\dfrac{1}{2}\frac{1}{\vert {\bf k}\vert}\;.
\end{equation}

The renormalization functions associated to the electron self-energy are defined as:
\begin{equation}
G_0^{-1}-\Sigma(k^0,{\bf k})=Z_\psi^{-1/2}(k^0,{\bf
k})[k^0\gamma^0-Z_v(k^0,{\bf k})v\bm{\gamma} \cdot {\bf k}],
\label{zelectron}
\end{equation}
The extra parameter $v$ appearing in (\ref{elprop}) in the graphene case has an associated renormalization function $Z_{v}$ which is a new feature characteristic of graphene. In the Lorentz invariant massless QED in any dimensions, only the electron wave function renormalization is associated to the the electron self energy.
The computation of the graph in Fig. \ref{primitive} (a) gives
\beq
\Sigma^{(1)}(\mathbf{k})\sim \frac{g}{4}v\bm{\gamma}\cdot\mathbf{k}
\log \frac{{\bf k}^2}{\Lambda^2},
\label{cutoff}
\eeq
where g is the coupling constant to be discussed later, and $\Lambda$ is a high energy cutoff. The logarithmic divergence can be fixed by redefining the fermion velocity $v$ which becomes, after finishing the renormalization procedure, energy-dependent.
The result is by now well known:
\beq
\frac{v_F(E)}{v_F(E_0)}=1-\frac{g}{16\pi} \log(E/E_0),
\label{oneloopveloc}
\eeq
which relates the value of the Fermi velocity at an energy $E$, with that at a reference energy $E_0$, assuming that the two energies are sufficiently closed. Notice that $E_0$ is not a cutoff of the order of
the bandwidth but any energy where the Fermi velocity
is determined by experiments. The corresponding diagram in massless QED(3) does not have a logarithmic divergence because the photon propagator has an extra inverse power of momentum. In QED(4) this diagram induces a wave function renormalization. It is interesting to note that the electron charge in the graphene model is not renormalized (after fixing  the velocity divergences, the photon self-energy is finite at all orders in perturbation theory)  but the velocity renormalization induces a renormalization of the graphene structure constant. Since according to eq. (\ref{oneloopveloc}) the Fermi velocity increases at decreasing energies, the effective coupling constant  $g\equiv e^2/4\pi v_F$ decreases in the infrared making perturbation theory more accurate.

\section{The coupling constant of Coulomb interaction and perturbation theory}
\label{alpha}
As mentioned in the introduction perturbation theory calculations of any observable quantity are plagued by  infinities. Perturbative renormalization \cite{C84} is a well prescribed mechanism to get rid of the infinities and define a sensible predictive theory. In the graphene system we encounter an additional problem: The bare value of the graphene fine structure constant
\beq
\alpha_G=\frac{e^2}{4\pi v_F},
\label{gbare}
\eeq
is not small.
Plugging--in the value of the bare electron charge $e$ and the Fermi velocity measured in nanotube experiments \cite{LB01} $v_F\sim c/300$, $\alpha_G$ is estimated to be of the order of $\alpha_G\sim 2.3 - 2.5$. 

\subsection{Screening in graphene}
The estimate of $\alpha$ described above precludes the use of perturbation theory. 
This problem is not so severe for samples on a substrate, where the substrate dielectric constant reduces the effective coupling constant:
 \beq
\alpha_G=\frac{e^2}{4\pi \varepsilon_G v_F}.
\label{gepsilon}
\eeq
where $\varepsilon_G$ includes intrinsic contributions and effects due to the environment in which graphene is immersed.
 A simple estimate gives 
 $\varepsilon_G = ( \varepsilon_{air} + \varepsilon_{subs} ) / 2 \sim 2-3$ for typical substrates. An interesting topic is the intrinsic screening due to the graphene layer itself. Internal excitations in three dimensional insulators lead to a finite dielectric constant, which reduces the effective charge of the mobile carriers. By extrapolating measurements of the excitation spectra in graphite, it has been proposed that similar effects lead to a large intrinsic dielectric constant in single layer graphene\cite{RU10}, $ \varepsilon_G\sim 13$.
 
 It is interesting to note that internal excitations in insulators, including electron-hole pairs, can be described as polarizable dipoles, which induce a field ${\cal E}_{ind} ( \vec{\bf q} ) \propto \left| \vec{\bf q} \right|^2 {\cal E}_{ext} ( \vec{\bf q} )$, where $\vec{\bf q}$ is the momentum of the external field. In a two dimensional system dipoles induce a field which tends to zero at small momenta, so that they do not contribute to the dielectric constant. The only internal screening processes in graphene which can modify the dielectric constant are the excitations near the Dirac energy. Their one loop contribution to the polarizability is finite and it can be calculated analytically
\begin{equation}
\Pi({\bf k}, \omega)=i\frac{e^2}{8}\frac{{\bf
k}^2}{\sqrt{v_F^2{\bf k}^2-\omega^2}}\;. \label{bubble}
\end{equation}
This one loop result is independent on the nature of the
interaction since only electron propagators appear in the
calculation. The resulting dielectric constant is $\epsilon = 1 + ( N \pi \alpha ) / 8$, where $N=4$ is the number of fermion flavors (see next section). The inclusion of low order ladder diagrams\cite{KUC08} does not change significantly this value.

An independent estimate of $\alpha$ in graphene has been obtained from measurements of the carrier-plasmon interaction in samples with a finite carrier concentration\cite{BSetal10}. The result, $\alpha \approx 2.2$ is consistent with the previous analysis.

\section{Renormalization of the Fermi velocity. 1/N expansion}
\label{Nexp}
The first perturbative renormalization results on the running of the coupling constant in graphene were obtained in ref. \onlinecite{GGV94} on the basis that, even if initially the graphene coupling constant is not small, it runs towards small values in the infrared where perturbation theory can be trusted. The logarithmic renormalization described in Sect. \ref{qed} is a one loop RG result that assumes the coupling constant to be small.

A standard perturbative framework for strongly coupled fermionic theories is the $1/N$ expansion first proposed in relation with Quantum Cromodynamics (QCD) \cite{H74}. $N$ is the number of fermionic species in the problem that in the case of graphene with the spin and valley degeneracies is $N=4$. This procedure was followed in the early publications \cite{GGV96,GGV99}  and was later retaken in \cite{FA08,DL09b,GGG10}.

The simplest non-perturbative calculation amounts to compute  the graph in fig. \ref{primitive} (a) substituting the photon propagator by the RPA result which is a resummation of the planar diagrams dominant  in the 1/N approximation:
\beq
{\cal P}(\omega, {\bf k})=\frac{-i}{2\vert{\bf k}\vert+
\frac{e^2}{8}\frac{k^2}{\sqrt{v_F^2 k^2 - \omega_k^2}}}.
\eeq
From that we get the following equation for the running Fermi velocity \cite{GGV99,FA08}:
\beq
\frac{1}{v_F}\frac{\partial v_F}{\partial l}=-\frac{8}{\pi^2 N}\big(1+\frac{\arccos g}{g\sqrt{1-g^2}}\big)
+\frac{4}{\pi g}, \label{large_n}
\eeq
where $g=\frac{Ne^2}{32 v_F}$ (the large N limit amounts to take the limit $N\to\infty$ keeping $g$ fixed). As we see the dependence on $g$ is non-perturbative and the growth of the velocity at low energies is slightly different than that in (\ref{oneloopveloc}).

Another interesting prediction done within this framework is the linear dependence of the quasiparticles lifetime  with the energy \cite{GGV96}, a distinctive of the marginal Fermi liquid behavior \cite{Vetal89}. Experimental evidences  of this behavior have  been described in \cite{Zetal06,Jetal07,LLA09,KLetal11}.

\section{Renormalization of the Fermi velocity: experimental confirmation}
\label{ren}
The deduction of the  values of the physical parameters from a given experimental
measure is one of the most important aspects of the renormalization program. This is
one of the issues where physics from different areas come into play, the best example being
the determination of the fine
structure constant of QED. By combining  solid state
measurements of the electron precession in magnetic fields or the quantum Hall effect,
with theoretical QFT calculations of the anomalous magnetic moment of the
electron we reach the actual precision of better than one part in a
trillion \cite{HFG08}.

It is fair to say that the Fermi velocity plays in graphene a role similar
to the effective mass in the standard Fermi liquid theory: it enters in the
definition of almost all the phenomenological quantities and, as such, it permeates
in the value of the experimental observations.  Taking for granted
a constant, absolute value for this parameter can misguide the interpretation
of some experiments or hide interesting physical issues.
There has been a very recent experimental report on the observation of the renormalization of the Fermi velocity \cite{EGetal11} confirming earlier more indirect evidences
\cite{BOetal07,Letal08}. This is a very important  result which can also affect the interpretation of other experimental observations. On the conceptual point of view, the verification of the growth of the Fermi velocity at low energies is interesting not only for the condensed matter community but also for the high energy as a beautiful example of renormalization at work. Under a more applied point of view, the Fermi velocity being almost the only parameter of the graphene model, it affects most of the experimental findings. The recognition that it changes with the energy
can offer new insights in the interpretation of infrared data.

\begin{figure}
\begin{center}
\includegraphics[scale=0.5]{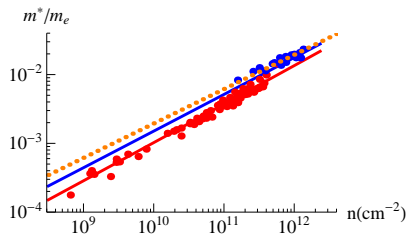}
\caption{Experimental measurements of the effective mass of carriers in high mobility graphene samples. Red dots give results in suspended samples, and blue dots are for samples on BN substrates. The red and blue lines are fits using eq. \ref{large_n}. Red: $\varepsilon_G = 1$. Blue: $\varepsilon_G = 4$. The dotted orange line shows a fit obtained neglecting the renormalization of the Fermi velocity. }
\label{expts}
\end{center}
\end{figure}
The renormalization predicted in eq.~\ref{large_n} is confirmed by the experimental results in\cite{EGetal11}. This reference reports measurements of the effective mass at different carrier densities for high mobility suspended graphene and graphene on BN. These experiments use the temperature dependence of Shubnikov-de Haas oscillations to infer the dependence of the Fermi energy, $\epsilon_F$, on the area of the Fermi surface, $S_F$. The effective mass is defined as
\begin{align}
m_{eff} &= \frac{\hbar^2}{2 \pi} \frac{d S_F}{d \epsilon_F}
\end{align}
so that for graphene
\begin{align}
m_{eff} &= \frac{\hbar k_F}{v_F}
\end{align}
A comparison of experimental results an fits based in eq.~\ref{large_n} are shown in Fig.~\ref{expts}.

\section{Opening a gap in graphene. Excitonic transition}
Graphene is a semimetal and has zero gap. Under the QFT point of view,
the electron mass (gap) is protected by the 3D version of chiral symmetry
and hence  a gap will not open by radiative corrections
at any order in perturbation theory. Quantum electrodynamics (QED) in (2+1) dimensions has been studied
at length in quantum field theory mostly in connection with
dynamical mass generation \cite{P84,AP81,JT81,N89} and confinement
\cite{Po75,Po77}. Both problems are crucial in the physics of
graphene, the first is related with the possibility of opening a
gap in the system, the second one with the screening of the
Coulomb interaction. A very
interesting review on the QED properties of graphene and its
possible dynamical mass generation  can be found in \cite{GSC07}.

As mentioned in Sect. \ref{qed}, the coupling constant of QED(3) has dimensions of energy
and, in this sense, mass generation is greatly facilitated since there is already a mass parameter in the theory. Even then, spontaneous chiral symmetry breaking in QED(3) in the 1/N approximation usually requires an unphysical number of fermionic species $N<2$ \cite{ABetal86}.

The situation with graphene is even worse since there is no mass parameter to begin with.
The simplest non-perturbative calculation that can be done to study the issue of the gap opening in graphene is the RPA type  described in Sect. \ref{Nexp} \cite{GGV99}. The absence of a constant term in the inverse electron propagator ensures that no mass is generated in this approximation. The 1/N expansion approach to the problem  has been revised recently in \cite{G10} with the conclusion that  a gap will not open for the physical values of the electronic degeneracy in graphene (N=4). A variational approach to the excitonic phase transition in graphene including the renormalization of the Fermi velocity \cite{SSG10} also produces quite negative results. A topological analysis of the stability of the Fermi points in graphene against gap opening
was done in \cite{MGV07}.

Despite these negative results, many reports on opening gaps in neutral graphene are found in the literature both on the experimental  and theoretical  sides. The possibility of an excitonic
insulator  was first suggested by Khveshchenko \cite{K01,K01b} which also raised the issue  of  magnetic catalysis \cite{Getal02} based on earlier studies of QED(2+1) \cite{FKM99}.

The possibility of opening a gap when the substrate is commensurate
with the lattice was  suggested in \cite{MGV07} and given as a
possible explanation of a gap observed in photoemission
experiments on epitaxially grown graphene\cite{GFetal07} and in
scanning tunneling microscopy experiments of graphene on graphite
\cite{LLA09}. The influence of the substrate in this context has
also been analyzed in \cite{R08}. Monte Carlo calculation in a lattice gauge theory framework
predict the opening of a gap \cite{DL09a,DL09b}. A gap has been also claimed in
very recent Angular resolved photoemission spectroscopy measurements of
epitaxial graphene upon dosing with small amounts of atomic
hydrogen \cite{BOetal09}. Disorder can give rise to various strong
coupling phases \cite{SGV05,FA08,HJV09} whose physical
properties only been partially  explored \cite{HJV08}.

The issue of the gap opening in single layer graphene remains open and some experimental effort
is needed to dillucidate it. The situation in the bilayer 	\cite{MF06,CNetal07} is more clear because the gap is not protected there, short range interactions can be active at low energies and gaps can open in various ways \cite{CNetal07,Oetal08,AC09,BY09,FMetal09,GGM10,Xetal10,XC10,WAetal10,NL10}.

\section{Retarded Coulomb interaction. Emergent Lorentz covariance}

As we have discussed, in the static model commonly used in graphene with an instantaneous photon propagator, the Fermi velocity grows without bound in the infrared \cite{GGV94}. This sets a lower bound on the validity of the model that breaks down when the Fermi velocity approaches the speed of light $c$.
The scale defines an infrared cutoff $\delta$ for the theory, and can be computed as:
\begin{equation}
\delta= k_R \exp\left[-\frac{16\pi(c-v( k_R))}{e^2}\right],
\end{equation}
where $k_R$, $v_R$ are the renormalization condition chosen  to define the  theory. A typical estimate
will of course produce an energy of approximately 100 orders of magnitude below $k_R$. This discussion is very similar to that leading to the Landau pole of QED \cite{BS80} and simply points to the incompleteness of the theory. In our case there are physical low energy bounds much less stringent than that, preventing
to access  the infrared region.  But even if this does not impose any real bound on the validity of the model from a experimental point of view, it is interesting to know that  a complete theory exists of which the static limit is only an approximation.

In the retarded model \cite{GGV94,GMP10} the Fermi velocity does not run without stop. The beta function of the Fermi velocity has  a non--trivial fixed point and the value of $v_F$ at the fixed point is precisely the speed of light $c$. This  is a very neat example of emergent Lorentz symmetry.

\section{Discussion}

The most important parameter in the computation of the observable quantities in graphene is the renormalized Fermi velocity that defines, together with the polarizability, the Coulomb coupling constant.

The program to fix this constant must be similar to the one followed in the case of the electromagnetic fine structure constant \cite{HFG08}. In the case of graphene the theoretical determination is much simpler since there are no mass parameters relations involved in the calculations. Moreover we have  the additional fact that the quantity determining $\alpha_G$ (Fermi velocity) is by itself an observable related to the one particle properties of the system. We expect that a proper combination of photoemission \cite{Zetal06,BOetal07} optical \cite{NBetal08} and transport \cite{LLA09,DSetal08,EGetal11} measures with the corresponding calculations in the QFT renormalization scheme  should be enough to determine $\alpha_G$ as precisely as needed both theoretically and experimentally, similarly to what happens in QED(3+1).

\section{Acknowledgments}
Support by MEC (Spain)
through grants FIS2008-00124, PIB2010BZ-00512 is acknowledged.

\bibliography{Renormalization}

\end{document}